\documentclass[journal, twocolumn]{article}
\usepackage{cite}
\usepackage{amsmath,amssymb,amsfonts}
\usepackage{algorithmic}
\usepackage{graphicx}
\usepackage{graphicx}

\usepackage{amsmath}
\usepackage{array}
\usepackage{caption}
\usepackage{floatrow}   
\usepackage{subcaption}
\usepackage{graphicx,xcolor} 
\usepackage[colorinlistoftodos]{todonotes}

\usepackage{breqn}
\usepackage[framemethod=tikz]{mdframed}
\usepackage[tableposition=top]{caption}
\usepackage{float}
\floatstyle{plaintop}
\restylefloat{table}
\usepackage{arydshln}
\usepackage{hyperref}
\usepackage{pifont}
\usepackage{amssymb}% http://ctan.org/pkg/amssymb
\usepackage{pifont}% http://ctan.org/pkg/pifont
\newcommand{\cmark}{\ding{51}}%
\newcommand{\xmark}{\ding{55}}%
\usepackage{ifpdf}
\usepackage{booktabs}
\usepackage{makecell}
\usepackage{sidecap}

\sidecaptionvpos{figure}{c}

\newcommand{\comment}[1]{}
\usepackage{ifpdf}

\usepackage{cite}

\begin{document}
\title{Ultrasound Scatterer Density Classification Using Convolutional Neural Networks by Exploiting Patch Statistics}
	%
	%
	% author names and IEEE memberships
	% note positions of commas and nonbreaking spaces ( ~ ) LaTeX will not break
	% a structure at a ~ so this keeps an author's name from being broken across
	% two lines.
	% use \thanks{} to gain access to the first footnote area
	% a separate \thanks must be used for each paragraph as LaTeX2e's \thanks
	% was not built to handle multiple paragraphs
	%
	
\author{Ali K. Z. Tehrani, Mina Amiri, Ivan M. Rosado-Mendez,\\ Timothy J. Hall, and Hassan Rivaz% <-this % stops a space
\thanks{\textcolor{red}{This work has been submitted to the IEEE for possible publication. Copyright may be transferred without notice, after which this version may no longer be accessible.} This work is supported  by  the  Natural  Sciences  and Engineering  Research  Council  of  Canada  (NSERC)  RGPIN-2020-04612.}% <-this % stops a space
\thanks{A. K. Z. Tehrani, M. Amiri and H. Rivaz are with the Department
	of Electrical and Computer Engineering, Concordia University, Canada.	
	Ivan M. Rosado-Mendez is with Universidad Nacional Autonoma de Mexico, Mexico.
	Timothy J. Hall is with the Department of Medical Physics, University of Wisconsin, United States.
	e-mail: A\_Kafaei@encs.concordia.ca, Amirim@encs.concordia.ca, irosado@fisica.unam.mx, tjhall@wisc.edu and 
	hrivaz@ece.concordia.ca}%		
		%and~Jane~Doe,~\IEEEmembership{Life~Fellow,~IEEE}% <-this % stops a space

%\A. K. Z. Tehrani and H. Rivaz are with the Department
%of Electrical and Computer Engineering, Concordia University, Canada,
%e-mail:A\_Kafaei@encs.concordia.ca and 
%hrivaz@ece.concordia.ca
		% <-this % stops a space
		
\thanks{Manuscript received; revised}}

\maketitle
	
	% As a general rule, do not put math, special symbols or citations
	% in the abstract or keywords.
\begin{abstract}
Quantitative ultrasound (QUS) can reveal crucial information on tissue properties such as scatterer density. If the scatterer density per resolution cell is above or below 10, the tissue is considered as fully developed speckle (FDS) or low-density scatterers (LDS), respectively. Conventionally, the scatterer density has been classified using estimated statistical parameters of the amplitude of backscattered echoes. However, if the patch size is small, the estimation is not accurate. These parameters are also highly dependent on imaging settings. In this paper, we propose a convolutional neural network (CNN) architecture for QUS, and train it using simulation data. We further improve the network performance by utilizing patch statistics as additional input channels. We evaluate the network using simulation data, experimental phantoms and \textit{in vivo} data. We also compare our proposed network with different classic and deep learning models, and demonstrate its superior performance in classification of tissues with different scatterer density values. The results also show that the proposed network is able to work with different imaging parameters with no need for a reference phantom. This work demonstrates the potential of CNNs in classifying scatterer density in ultrasound images.
\end{abstract}
	
	% Note that keywords are not normally used for peerreview papers.
%\begin{keyword}
%Quantitative Ultrasound, Scatterer density, Convolutional Neural Network, Deep learning
%\end{keyword}

	% For peer review papers, you can put extra information on the cover
	% page as needed:
	% \ifCLASSOPTIONpeerreview
	% \begin{center} \bfseries EDICS Category: 3-BBND \end{center}
	% \fi
	%
	% For peerreview papers, this IEEEtran command inserts a page break and
	% creates the second title. It will be ignored for other modes.
%\IEEEpeerreviewmaketitle

\section{Introduction}
Ultrasound imaging is increasingly attracting the attention of researchers and clinicians due to being a real-time and non-ionizing imaging modality, and being less expensive and more portable compared to other medical imaging techniques. However, several types of artifacts make interpretation of ultrasound images difficult. Cells, collagen, microcalcifications, and other microstuctural components are often smaller than the wavelength of the ultrasound wave, and scatter the wave and create the granular appearance called speckles. The scattered signal from scatterers provides useful information about characteristics of the scatterers which are highly related to the tissue properties. Quantitative ultrasound (QUS) measures the tissue characteristics by analysing the ultrasound signal \cite{Dutt1995,NAM2011,Insana1990,mamou2013quantitative,Vajihi2018,rouyer2016vivo}. It aims to provide quantitative estimations of tissue characteristics which cannot be otherwise obtained from the B-mode image. It has been employed in many different applications such as liver ﬁbrosis \cite{ho2012, Pirmoazen2020}, bone quality measurement \cite{chin2013}, breast tumor classiﬁcation \cite{Shankar2001,LARRUE2014} and cardiac tissue characterization \cite{CLIFFORD1993}. Improving QUS techniques can eventually broaden the applications of this safe and cost-effective method in diagnosis and treatment of a large number of disorders.\par
 QUS methods can be classified into two broad categories: spectral-based and envelope-based methods \cite{Oelze2016}. Parameters such as the backscatter coefficient and attenuation coefficient can be estimated by spectral-based methods, usually with a requirement of a reference phantom to remove system-dependent effects\cite{ yao1990backscatter,Vajihi2018,deeba2019spatially}. In envelope-based methods, different characteristics of the tissue is usually estimated by analysing and modeling the envelope of the ultrasound RadioFrequency (RF) data by fitting a probability density function. The sample size, wave frequency, and the attenuation can affect the accuracy of the distribution modeling, and therefore its parameter estimations \cite{Cloutier2004,tsui2017effect}.\par

 The statistics of echo-envelope data, extracted by either model-based or model-free parameters, provide information about tissue properties. Model-based parameters try to fit a distribution to the envelope data. If there are many scatterers (more than 10 in a resolution cell (an ellipsoidal volume defined by - 6 dB point of the beam profile\cite{wagner1983statistics})), the envelope data is considered as a fully developed speckle (FDS), and the RF data can be modeled by the Gaussian distribution; therefore, envelope values follow the Rayleigh distribution \cite{Dutt1995,Rosado2016,Rivaz2006}. However, when the number of scatterers is low, the resolution cell has low-density scatterers (LDS); therefore, the Rayleigh distribution fails to model the envelope statistics. To model LDS, other distribution such as K-distribution\cite{CLIFFORD1993}, Homodyned K-distribution\cite{prager2002,Destrempes2013} and Nakagami distribution\cite{Shankar2001} can be utilized. Among these, Homodyned K-distribution is the most comprehensive but the most complex that does not have a closed-form solution. The Nakagami distribution provides a good estimate of the envelope signal with low-complexity and is widely used in QUS studies.\par 
 The Nakagami image, originally proposed by \cite{Shankar2001}, can be used to describe different probability density functions, and to characterize various scatterer patterns in tissues. it has been shown to be useful in discriminating different scatterer concentrations and tissue types. The Nakagami image can depict tissue properties that are not visible in ultrasound B-mode images, and has been employed in several studies for tissue characterization \cite{tsui2008,tsui20082,tsui2011,ho2012}.\par 
 
Model-free parameters such as the envelope signal to noise ratio (SNR), skewness (S) and entropy \cite{Tsui2017,Rivaz2006} are statistical parameters that change with different scatterer distributions.
Entropy parametric imaging is a QUS imaging technique, which uses a small sliding window throughout the image to measure the entropy (the overall level of variations) of the backscattered RF signal. It has been shown to be effective in differentiating tissues with different scattering properties, and can provide higher accuracy in a smaller patch size compared to Nakagami imaging \cite{Tsui2017}.\par

Deep Learning (DL) techniques have been utilised in many fields of medical image processing. They have also proved useful in different ultrasound applications such as segmentation\cite{Amiri2020, meiburger2018automated}, super resolution imaging\cite{van2019deep,brown2020deep,Goodarzi2019} and elastography\cite{gao2019learning,evain2020pilot, Kafaei2020}. A few studies have also attempted to tackle the challenge of extracting quantitative measures from ultrasound images using DL techniques. Byra \textit{et al.} \cite{Byra2017} used Nakagami images to train a convolutional neural network for the task of breast lesion classification. Wang \textit{et al.} \cite{Wang2020} have proposed a 3D convolutional network for breast cancer detection. However, the appearance and even statistics of ultrasound images can change with changes in imaging parameters such as time gain compensation and focal points. Such changes are well studied in DL and are referred to as domain shift \cite{stacke2019closer}. If not accounted for, domain shift renders DL estimates grossly inaccurate. In fact, this is one of the reasons that DL is less explored in QUS compared to other ultrasound applications. \par 
In a recent work, we designed a CNN to classify FDS and LDS \cite{Kafaei20202}. The CNN was fed with envelope data and the RF data spectrum from small patches of RF data, and  was compared with a Multi-Layer Perceptron (MLP) classifier, which used SNR and skewness as inputs. We used patches to analyse a small area of the image and therefore, to provide a high resolution. The effect of patch size was also investigated (with patches sized 5 to 10 $\times$ wavelength). The results showed that in small patch sizes, the CNN outperformed the MLP classifier, whereas for larger patch sizes, where the statistics of the patch could be reliably estimated, the MLP classifier outperformed the CNN.\par
In another recent work, we segmented simulated images with three different scatterer densities using a U-Net \cite{Amiri20202}. We found that the network was able to segment precisely when the intensity difference between the inclusion and the background was high thus the network could associate the intensity to the scatterer density.\par
In \cite{zhang2020deep}, the mean scattering cross section which is another QUS parameter was estimated for the whole image. They assumed that all regions have FDS which is a limiting factor in real ultrasound images. 
In this study, we aim to classify FDS from LDS regions using CNNs in small patches (Note that the patch size is different for simulation, experimental and \textit{in vivo} data) where classical statistical parameters commonly used in QUS studies cannot be estimated accurately. We use the ultrasound envelope data as the input to the network, since statistics such as SNR and Nakagami parameters are histogram-based, meaning that they ignore image texture. We hypothesize that the texture of the ultrasound envelope image contains crucial information which can be useful to determine the density of scatterers. \par
We use a large amount of simulated data to train the network, and test the network on simulated, phantom and \textit{in vivo} data. We show the effect of domain shift \cite{stacke2019closer} caused by changes in imaging parameters on the results and the feature space. We also demonstrate that using both texture information and patch statistics helps improve the robustness against imaging parameters. We validate our methods using experimental phantoms and \textit{in vivo} data which have different imaging parameters compared to the data used for training the networks. We also show the effect of transfer learning for domain adaptation. Our contributions can be summarized as follows:
\begin{itemize}
  \item A CNN architecture is proposed to analyse the ultrasound envelope data for tissue characterization.
  \item The network is further improved by exploiting patch statistics which provide additional information. 
  \item Extensive ablation experiments are done to find the best input for the network in simulation and experimental phantoms, and to improve the robustness against variations of imaging parameters.
 \item Transfer learning is investigated to evaluate the network's performance when it is fine-tuned with images of the test domain. 
\item Different classifiers (support vector machine (SVM), random forest and MLP) are used to classify based on patch statistics.
  \item Experimental phantom and \textit{in vivo} data are employed to validate our work in different imaging settings.
\end{itemize}

\section{Methods}
In this section, we first describe different datasets we analysed. We then present the scatterer density classification methods developed in this work, which include both classical (SVM, random forest and MLP) and DL methods (CNN and CNN with patch statistics as additional inputs), and provide intuitions for using different inputs. 
\subsection{Data}
We employed 3 different datasets to investigate the performance of our proposed methods as outlined below.
\subsubsection{Simulation data}
We simulated 200 phantoms of size 30 mm$\times$ 30 mm$\times$ 1 mm using the Field II pro toolbox \cite{Jensen1997}, with the center frequency of 6.67 MHz. The sampling frequency was 100 MHz and the RF signals were then downsampled to 50 MHz. Other imaging parameters can be found in Supplementary Materials.\par
We randomly distributed point scatteres in the phantoms. In 100 FDS phantoms, we included 16 scatterers per resolution cell. In the remaining 100 LDS phantoms, we included 2 scatterers per resolution cell. The resolution cell size was determined by calculating the correlation between the data and a moving window in different regions \cite{rivaz20079c}. The size was 0.1 $mm^2$ at the focal point (The out of plane resolution cell size is not computed). 
We randomly cropped 5000 patches of size 256$\times$32 (4.04 $mm$ $\times$ 5 $mm$ which is 17 and 21 $\times$ wavelength in axial and lateral directions, respectively) from different depths as the training set and 1000 patches as the validation set. For the test set, we simulated 20 more phantoms with a random scatterer density value of 2 or 16 $\pm$ 10\% in order to make the test data more challenging. We randomly selected 500 patches from these phantoms as the test set to evaluate the methods. This dataset will be publicly available online after acceptance of this paper at data.sonography.ai.\par

\subsubsection{Experimental phantom}
Three different phantoms were used to validate our method.
 The phantoms were of size 15cm$\times$ 5cm$\times$ 15cm, built from homogeneous mixture of agarose gel media and glass beads as scattering agents. 
The glass bead diameter range and bead concentration in the phantoms are reported in Table \ref{table1}. For more information on construction details, the speed of sound and attenuation coefficient of these phantoms refer to \cite{rosado2014}. 
 The phantoms were imaged using an 18L6 transducer operating at 10 MHz frequency using an Acuson S2000 scanner (Siemens Medical Solutions, Malvern, PA) and the RF data was acquired using Axius Direct Ultrasound Research Interface \cite{brunke2007ultrasound}. We computed the resolution cell size using correlation method at different depths and it varied between 0.284 $mm^3$ (at the top where resolution was poor) and 0.036 $mm^3$ (at the focal point where resolution was the highest). This high variation of the resolution cell size can have an adverse effect on the classification, especially when this variation is not observed by the network during training. The numbers of scatterers per resolution cell for different depth are given in Table \ref{table1}.\par
We used the experimental phantoms as the test data to evaluate the performance of different models optimized or trained on the simulation data (with or without fine-tuning the network). Phantom A (high concentration) belongs to the FDS class and Phantoms B (medium concentration) and C (low concentration) belong to the LDS class.

\begin{table*}[]
\centering
\caption{Characteristics of the experimental phantoms and their scatterer concentration per resolution cell using 18L6 transducer (the range shows the minimum and maximum values derived from different depths).}
\label{table1}
\begin{tabular}{@{}cccc@{}}
Phantom & \begin{tabular}[c]{@{}c@{}}Diameter of Random \\ Scatterers ($\mu m$)\end{tabular} & \begin{tabular}[c]{@{}c@{}}Scatterers Concentration \\ per $mm^3$\end{tabular}&\begin{tabular}[c]{@{}c@{}}Scatterers Concentration \\ per resolution cell\end{tabular} \\ \hline
A (High) & 5-40 & 236 & 8.50-67 \\
B (Medium) & 75-90 & 9 &0.32-2.55\\
C (Low) & 126-151 & 3&0.11-0.85
\end{tabular}
\end{table*}

\subsubsection{\textit{In vivo} data}
We used breast ultrasound images recorded by a Siemens Sonoline Elegra System (Issaquah, WA) with the sampling frequency of 36 MHz and a lateral beam spacing of 200 $\mu m$. I-Q echo data were recorded in a file on the imaging system when data acquisition  was stopped (frozen on the imaging system) \cite{brunke2007ultrasound}. The I-Q data were converted to RF echo data offline using the known modulation frequency of the imaging system. More information about this dataset and the recording procedure is provided in \cite{zhu2002modified}. 
\subsection{Classical Statistical Parameters}
Several parameters have been proposed in literature for estimating the scatterer density in ultrasound images. SNR and skewness are among the most important parameters proposed to classify different scatterer densities:
\begin{equation}
\begin{gathered}
R = SNR=\frac{\overline{A^v}}{\sqrt{\overline{A^{2v}}-(\overline{A^v})^2}},\\ \\
S=skewness= \frac{\overline{(A^v-\overline{A^v})^3}}{(\overline{A^{2v}}-(\overline{A^v})^2)^{1.5}}
\end{gathered}
\end{equation}
where $A$ is the envelope of RF data, $v$ is the signal power and $\overline{(...)}$ denotes mean operation. While in \cite{Dutt1995}, $v$ smaller than 1 was suggested due to having higher dynamic range and lower estimation error, Prager \textit{et al.} proposed 1.8 as the optimal value \cite{prager2001speckle} in terms of the estimation error. We analysed both recommended values of 0.5 and 1.8, and obtained significantly better results on the validation set using the 0.5 value (Area Under Curve (AUC) of 0.894 vs. 0.876 when employing the MLP, and 0.802 vs. 0.794 when employing the SVM classifier). We, therefore, set $v$ to 0.5 in this study.\par
When the patch size is big enough, the estimation error of $R$ and $S$, and therefore the classification error based on these parameters is low. But for small-size patches, the classification becomes difficult \cite{Dutt1995,Kafaei20202}. This is especially important in clinical applications where tissues are rarely homogeneous and a large patch may include different scattering properties~\cite{Rivaz2006,prager2002}.\par
Another parameter which has been proposed for scatterer density classification is entropy \cite{Tsui2017}:
\begin{equation}
Entropy = \sum_{n=1}^{N} p(i) log[p(i)]
\end{equation}
where statistical histogram of the envelope data square is represented by $p$, and $N$ is the number of bins for calculating the histogram, which is arbitrarily set to 100 in this study. Entropy increases as the density of scatterers increases (moving from LDS to FDS). The entropy measure is shown to be effective when using a small window for QUS analysis \cite{Tsui2017}.\par
Another parameter that has been shown useful in estimation of scatterer density is the Nakagami model parameters $m$ (a maximum likelihood estimator of the shape parameter) and $T$ (a generalized likelihood ratio test statistic) \cite{Rosado2016}:
\begin{equation}
\begin{gathered}
m = \frac{(\overline{A^2})^2}{var[A^2]},\\ \\
T=2K(log\frac{m^m}{\Gamma(m)} + (m-1)[\overline{log(I)}-log(\overline{I})-1])
\end{gathered}
\end{equation}
where $A$ is the envelope data and $\Gamma$ represents the Gamma function. $I$ is a vector representing $K$ independent and identically distributed samples of the intensity from a specific patch. Different values of $m$ explain different properties; when the $m$ parameter approaches 1, the distribution approaches the Rayleigh distribution. The $m$ parameter above and below 1 represent post- and pre-Rayleigh distributions, respectively, which are forms of a more general family of distributions, called Rician \cite{tuthill1988,Shankar2001,Dutt1995,wagner1983statistics}. \par
There is a strong correlation between features $m$ and $T$. The features $m$ and $R$ are also highly correlated. We therefore, remove the feature $m$ from the feature list to eliminate the redundancy between different features. We consider aforementioned parameters as a set of features to classify FDS and LDS patches. The Supplementary Materials contain detailed correlation analysis of these statistical parameters.\par
\begin{figure}
		\centering
		\begin{subfigure}{0.47\linewidth}
			\centering
			\includegraphics[width=0.18\textheight]{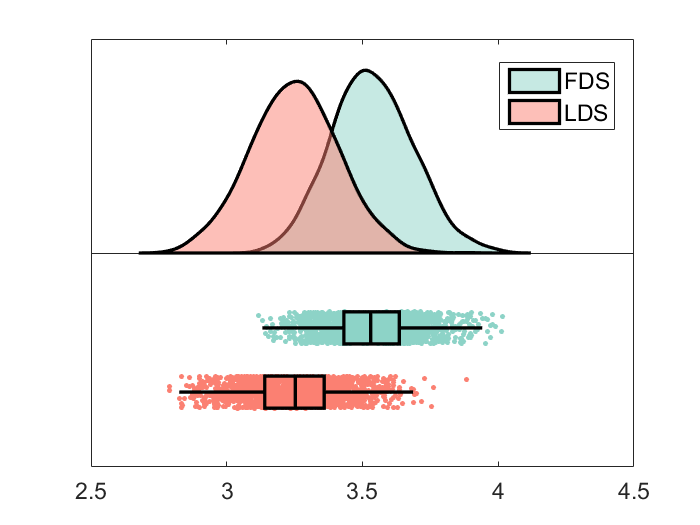}
			\caption{R}
		\end{subfigure}
		\begin{subfigure}{0.47\linewidth}
			\centering
			\includegraphics[width=0.18\textheight]{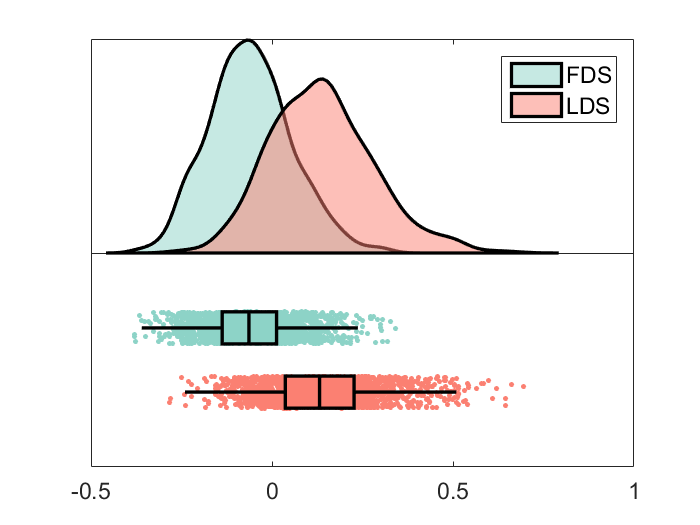}
			\caption{S}
		\end{subfigure}
		\begin{subfigure}{0.47\linewidth}
			\centering
			\includegraphics[width=0.18\textheight]{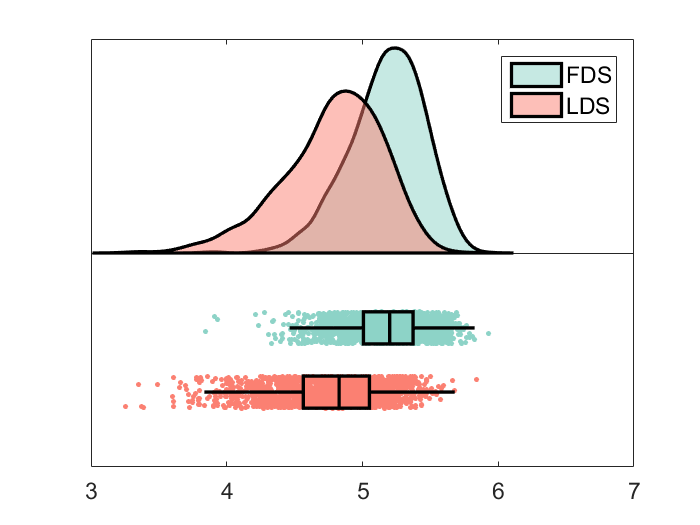}
			\caption{Entropy}
		\end{subfigure}
			\begin{subfigure}{0.47\linewidth}
			\centering
			\includegraphics[width=0.18\textheight]{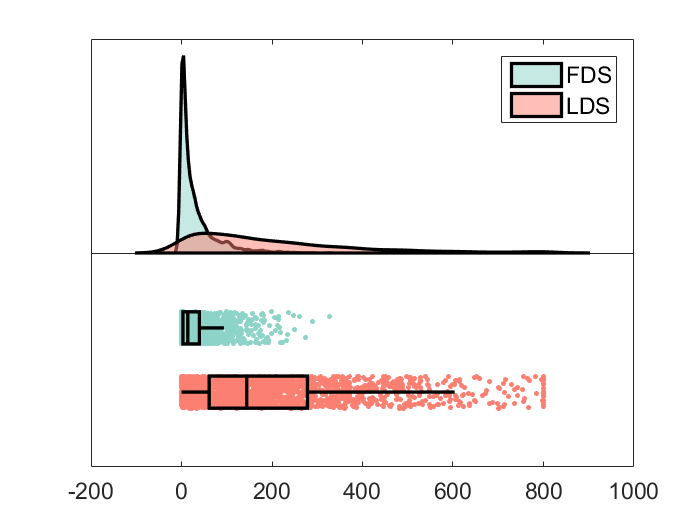}
			\caption{T}
		\end{subfigure}
		\caption{The distribution of the patch statistics for FDS and LDS in simulated training data. The patch size is small enough such that FDS and LDS classes overlap.}
		\label{figure1}
			\end{figure}

Fig. \ref{figure1} shows the distribution of different features, extracted from the simulation training data for LDS and FDS classes. The patch size is small so that for all features, a considerable overlap exists between the distributions of the two classes, which makes the classification highly erroneous using only a single feature. As opposed to our previous work \cite{Kafaei20202} where only parameters $R$ and $S$ were used for classification, we use $R$, $S$, entropy and $T$ together to obtain higher performance in classification. As shown in Fig. \ref{figure1}, the dynamic range of the features are not similar, hence we employ normalization (they are normalized to be in range 0-1) across each feature in the training data. The test and validation data were also normalized using the same coefficient obtained from the training data.
\subsection{Machine Learning Methods}
In order to classify FDS and LDS classes, we developed classical machine learning techniques in addition to DL methods. In this section, we describe the details of these classic techniques.
\subsubsection{Support Vector Machine (SVM)}
We used SVM as a classical machine learning algorithm to classify FDS versus LDS. We analysed different SVMs with linear and non-linear (Radial Basic Function (RBF) and polynomial) kernels. An SVM with an RBF kernel led to the best results on the validation set, and was selected throughout this manuscript. 
\subsubsection{Random forest classifier}
Random forest is a learning method based on the decision tree algorithm and ensemble of different trees' outputs, and is amongst the top classification algorithms. By changing different parameters of a random forest model, we found the best performing model on the validation set, and used this model to classify different patches of simulation and experimental phantom data. 
\subsection{Deep Learning Methods}
\subsubsection{Multi-Layer Perceptron (MLP)}
We proposed an MLP structure to classify FDS and LDS groups. To find the best network architecture for classifying scatterer density using the aforementioned features, we investigated the performance of different MLP architectures on the validation data. We obtained the best results with a 3-layer network. Further increase in the number of layers did not improve the results and lead to overfitting, a common problem with MLPs. We also analysed different numbers of neurons in each hidden layer. We incrementally increased the number of neurons in two hidden layers. Including 128 neurons in the first hidden layer, and 32 neurons in the second hidden layer led to the best result. However, it is important to note that the results reached a plateau and did not change substantially by changing  the number of neurons. We employed Batch Normalization \cite{ioffe2015} in the first and second layers and Dropout \cite{srivastava2014} in the second layer. The activation functions were Tanh for the first two layers and Sigmoid for the last layer. The loss function was binary cross entropy and the network was optimized using the Adam optimizer.
\subsubsection{Convolutional Neural Network (CNN)}
We used a CNN to classify the scatterer density. Our proposed CNN structure is presented in Fig. \ref{figure2}. The input can have one or two channels. The FDS group was considered as the class with label 1, and the LDS group was considered as the class with label 0. The output of the network is the probability of being FDS. The network contains the following components: 
\begin{itemize}
  \item Residual blocks: Each residual block consists of 2 convolutional layers with a skip connection followed by another convolutional layer and an average pooling layer of size 2$\times$2. Using residual blocks provides the ability to train deeper and therefore more efficient networks by eliminating the vanishing gradient issue which happens in deep networks \cite{he2016deep}. 
  \item Convolution block: The network consists of two convolutional blocks before and after the residual blocks to extract features.
  \item Fully-connected layers: Two fully connected layers are used in the final step for classification. The first fully connected layer employs Dropout for regularization. \par
\end{itemize}
\subsubsection{CNN with patch statistics as additional inputs}
\begin{figure*}
		\centering
			\includegraphics[width=0.7\textheight]{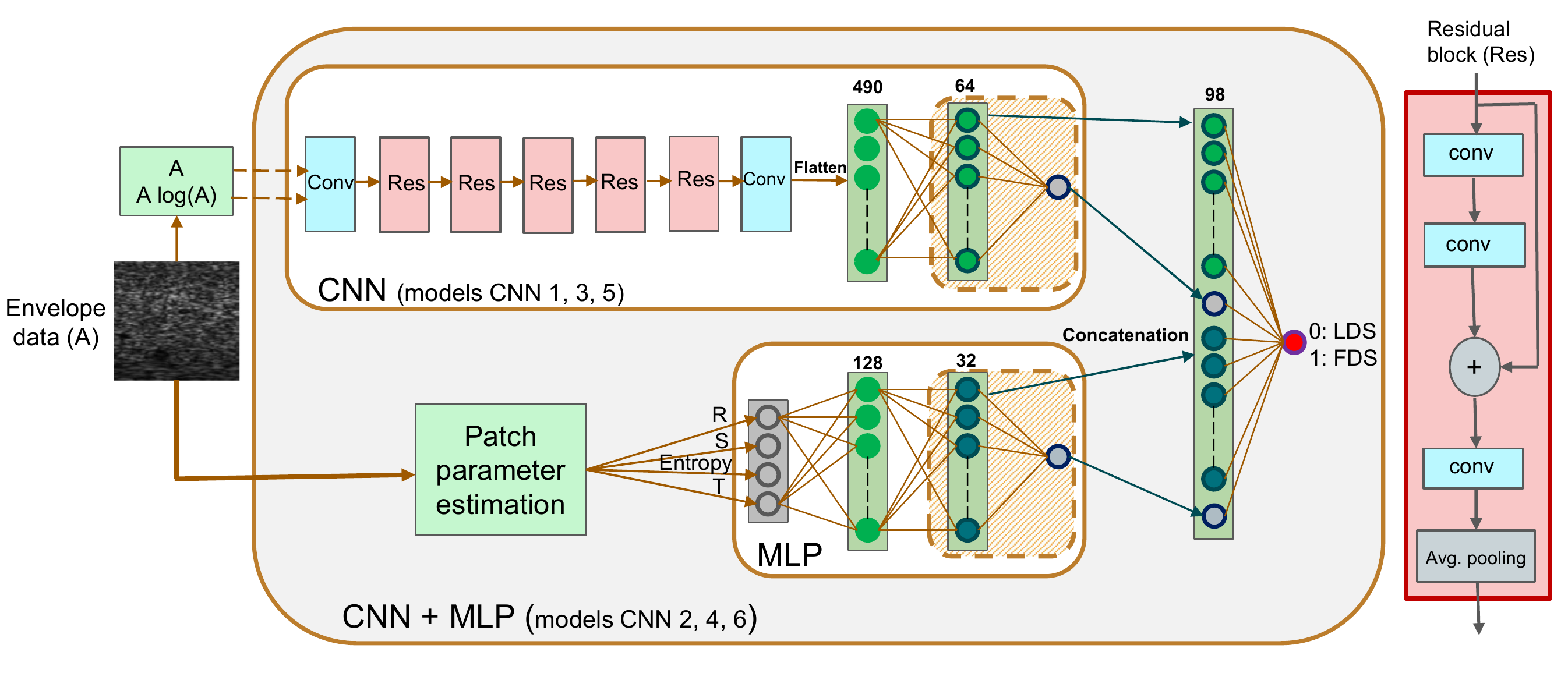}
			%\vspace{-1.5cm}%			
			\caption{Proposed architecture for different networks.}
			\label{figure2}
\end{figure*}

To further enhance the network, we propose to utilize the patch statistics ($R$, $S$, entropy and $T$) as additional inputs. We tested different settings to determine the optimal way to fuse the information of patch statistics to the CNN. Fig. \ref{figure2} shows the outcome ($A$ denotes envelope). The CNN part is the same as the CNN network described in section D-2 and the patch statistic classifier part is similar to the MLP explained in section D-1. These parameters are fed to an MLP to generate a feature map, which is concatenated to the feature map obtained from the CNN. The resulted feature map is then used for a final classification. \par
Our first intuition was to train the whole network end to end. However, the CNN and the MLP have vastly different numbers of parameters and this resulted in a low generalization and a large sensitivity to the initial seeds. To mitigate the imbalanced number of parameters, we proposed training each part separately. We then trained the fusing part while the CNN and MLP weights were kept fixed. Please refer to the Supplementary Materials for more information regarding the training strategy.

%\begin{figure*}
%	\centering
%	
%			\centering
%			\includegraphics[width=0.5\textheight]{net2.pdf}
%
%
%\end{figure*}
\subsection{Training Schedule}
To augment the data, random Gaussian noise, elastic deformation and random flipping in lateral direction were employed. The networks were trained with the Adam optimizer and the binary cross entropy was used as the loss function. Due to the fact that there were different networks with different inputs and to have a good generalization, we adopted a variant of early stopping which could be considered as a form of implicit regularization \cite{zhang2016understanding}. For early stopping, the validation AUC was selected as the stopping criteria; when the best validation AUC was reached during the training and remained the best after 20 epochs, we stopped the training. The cyclic learning rate was also used in order to avoid bad local minima \cite{smith2017cyclical}.
\subsection{Input Channels}
In \cite{Destrempes2013}, log compression of envelope along with the envelope have been used ($log(A)$ and $A^2\times log(A^2)$) for estimating statistics using the Homodyned K-distribution. Inspired by their work, we used $A\times log(A)$ as a novel input to the proposed CNNs. We therefore investigated different types of inputs for our proposed network architectures, including the amplitude alone $A$, $A\times log(A)$, and both $A$ and $A\times log(A)$ as two different channels (Fig. \ref{figure2}). We also tested other inputs including $log(A)$ and $\sqrt{A}$ but did not observe any improvements. For brevity, the results are not included in the manuscript.    
\subsection{Transfer Learning}
\label{fine_sec}
To evaluate the performance of our proposed networks on the experimental phantom data and \textit{in vivo} data, we performed transfer learning for the top performing model using 8 frames of phantoms A and B. The frames' location was far from the test frames to avoid data leakage. We randomly cropped 1000 patches and fine-tuned CNN and MLP separately. Similar to the training, we used cycling learning but with smaller learning rate since only small changes were required for the network adaptation.

\begin{table*}
\centering
\begin{tabular}{@{}ccccc@{}}
Model          & Input                                           &  Patch Statistics        & \textbf{AUC (\%)} & \textbf{Youdens' Index} \\ 
\hline

\hline
SVM & $A$                                      & \cmark &  0.892 (0.867- 0.920)  & 0.62 \\ 
\hline
Random forest & $A$                                     &  \cmark                  & 0.894 (0.868- 0.919)  & 0.62 \\ 
\hline
MLP & $A$                                      & \cmark                  & 0.894 (0.874- 0.912)  &0.61  \\ 
\hline
CNN 1              & $A$                                      & \xmark                 & 0.957 (0.942- 0.966) &0.79   \\ 
\hline
CNN 2              & $A$                                      & \cmark                 & 0.957 (0.944- 0.967) & 0.80   \\ 
\hline
CNN 3              & $A \times log(A)$        & \xmark         & 0.907 (0.887- 0.925) & 0.67           \\ 
\hline
CNN 4              & $A \times log(A)$         & \cmark        & 0.924 (0.904- 0.939)  & 0.71         \\ 
\hline
CNN 5              & $A$, $A \times log(A)$  & \xmark         & 0.963 (0.949- 0.973) &\textbf{0.82}         \\ 
\hline
CNN 6              & $A$,~$A \times log(A)$  & \cmark             & \textbf{0.964 (0.952-0.973)}&\textbf{0.82}              
\end{tabular}

\caption{Results of different classification models on simulation data.}
\label{table2}
\end{table*}
 \subsection{Evaluation Metrics}
To evaluate the classification performance, we used AUC of the Receiver Operating characteristic Curve (ROC) and also Youden's Index \cite{youden1950}. We estimated the 95\% confidence interval of AUC by employing boot strapping (i.e. sampling the data with replacement, for 1000 times).
Youden's Index is a measure of both sensitivity and specificity:  
\begin{equation}
\begin{split}
J=\frac{TP}{TP+FN}+\frac{TN}{TN+FP}-1 \\
= Sensitivity + Specificity - 1
\end{split}
\end{equation}
where $TP$, $FN$, $TN$ and $FP$ denote true positive, false negative, true negative and false positive, respectively.
We calculated the Youden's Index for different threshold values and the highest values are compared between different models.

\section{Results}
In this section, we provide the results of the proposed models for classification of FDS and LDS classes when analyzing different datasets. We tested three classifiers without including any CNN (i.e. a SVM, a random forest and an MLP model). Different CNNs were also analysed, by including (CNNs 2,4,6) or excluding the MLP branch (CNNs 1,3,5), and by considering different inputs to the network including the envelope alone (CNNs 1,2), the envelope multiplied by the log compressed envelope (CNNs 3,4), and both images as two separate channels (CNNs 5,6- Fig. \ref{figure2}). All DL models were trained according to the training schedule explained in Section II-E. The weights of the top-performing network will be publicly available online after acceptance of this paper at data.sonography.ai.

\subsection{Simulation Results}
\begin{figure}
		\centering
			\includegraphics[scale=0.34]{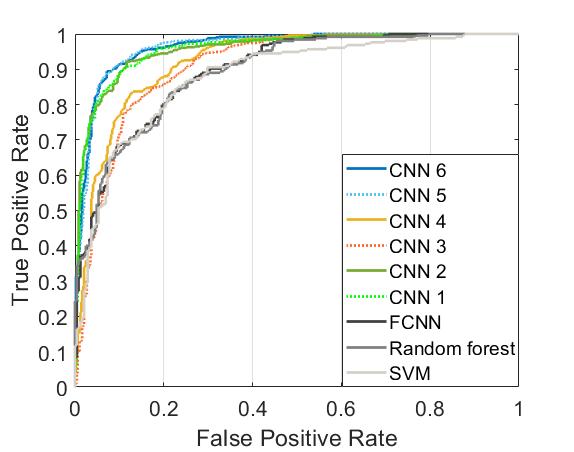}
		\caption{ROC curves of different models on simulation test data.}
		\label{figure_roc_sim}  	
	\end{figure}

All proposed models were evaluated on the simulation data. The AUC values and Youden's indexes are represented in Table \ref{table2}, and the ROC curves of different methods are shown in Fig. \ref{figure_roc_sim}. As seen in Table \ref{table2}, CNN-based models provide better results compared to the MLP and SVM and random forest models. CNN 6 achieves the best AUC (0.964) among all studied models.\par 
According to the results,  exploiting the patch statistics is helpful (compare CNN 1 vs. 2, 3 vs. 4 and 5 vs. 6). It shows that classical patch statistic parameters provide complementary information to the CNN. Interestingly, the envelope ($A$) gives better results compared to ($A \times log(A)$). However, using both inputs as two separate channels provides the best results (compare CNNs 1, 3 and 5).

\subsection{Experimental Phantom Results}
\begin{figure}
			\centering
			\includegraphics[scale=0.34]{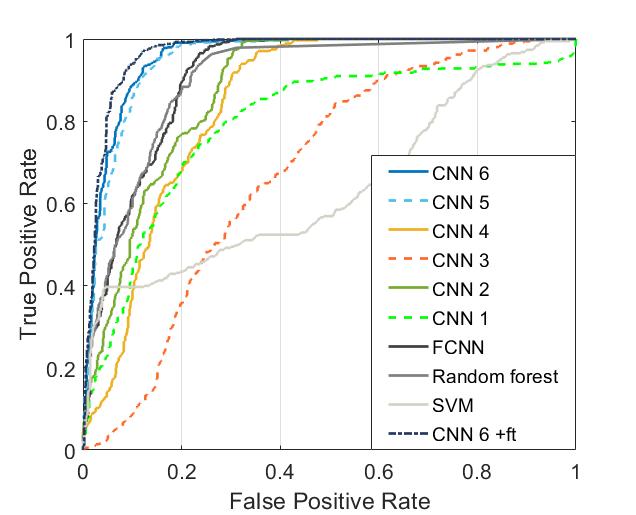}
			\caption{ROC curves of experimental phantoms. There is a considerable difference between single channel CNNs with/without utilizing patch statistics. CNN 5 and CNN 6 which use $A\times log(A)$ and $A$ as input channels perform well for experimental data. Fine-tuning further improve CNN 6 by removing domain shift.}
				\label{figure_roc_exp}  
		\end{figure}

The results of classifying small patches from phantom A vs. phantoms B and C are provided in Table \ref{table3}, and the ROC curves are shown in Fig. \ref{figure_roc_exp}. The patch size in terms of number of pixels is the same as the simulation data but it differs in terms of size in mm (4.92 mm$\times$ 4.28 mm). Similar to the results on the simulation data, utilizing the patch statistics (CNNs 5 and 6) boosts the results (Table \ref{table3} and Fig. \ref{figure_roc_exp}).\par

By comparing the results of simulation and experimental phantoms, we observe that the models with the envelope input (CNNs 1 and 2) perform well for the simulation data but poorly for the experimental data (compare Fig. \ref{figure_roc_sim} and Fig. \ref{figure_roc_exp}). The models with both inputs (CNNs 5 and 6) have the best results for both simulation and experimental phantom data. CNN 6 has the best result among all models without fine-tuning (AUC=0.955 and Youden's Index=0.79).\par

With only using the simulation data for training, we can still classify the experimental phantoms although the changes in imaging parameters and real artifacts in the experimental phantom data make the classification more challenging. However, employing both patch statistics and texture information extracted by CNNs using the proposed inputs ($A$ and $A\times log(A)$) improves the robustness of the network.\par

\begin{figure*}
		\centering
			\includegraphics[scale=.7]{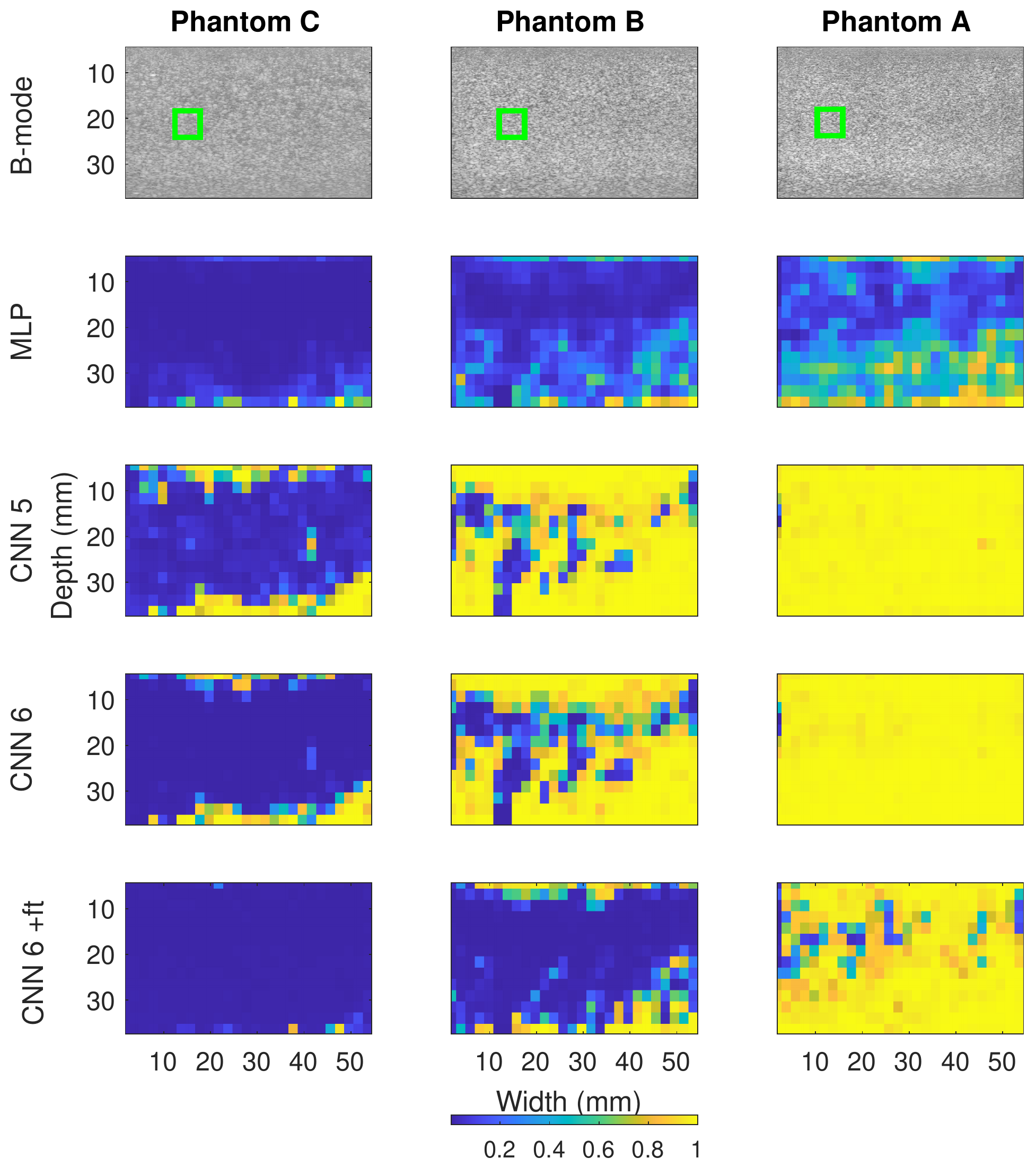}
			\caption{The results of MLP, CNN 5, CNN 6 and CNN 6 + ft models on the experimental phantoms. The color code represents the predicted output of the networks, from 0 (LDS) to 1 (FDS).}
			\label{figure4}
\end{figure*}

Fig. \ref{figure4} depicts some examples of the studied images using different models. We split each image into overlapping patches (50\% overlap), and feed all patches to the network. As seen in Fig. \ref{figure4}, CNNs 5 and 6 perform very well in classification of patches from phantoms A and C but they estimate high probability of being FDS for patches from phantom B which is not the correct class. However, according to the Table \ref{table3}, these two networks have very high AUCs (0.949 and 0.955). These apparent contradictory results can be explained by the fact that AUC is not sensitive to threshold values and by considering a high threshold for classification, phantom A can be discriminated from phantoms B and C. Because of the domain shift from the simulation to experimental data, the networks trained on the simulation data predict high probability of being FDS for the phantom B. Interestingly, the opposite effect is observed for the MLP which uses only patch statistics; The MLP results in low score values for all three phantoms which is in contrast to CNN 5. Combining both CNN and MLP (CNN 6) leads to an improvement. To tackle the domain shift problem, CNN 6 is fine-tuned with images of the experimental phantom domain. The fine-tuned network correctly predicts low probability of FDS for most of patches in the phantom B in contrast to the CNNs trained only on the simulation data. Please refer to the Supplementary Materials for further details.\par
We should also mention that the networks estimate higher probability of FDS in the top and bottom regions of the phantoms. The main reason is that in these regions, the resolution cell size is larger that focal region, leading to a larger number of scatterers per resolution cell.

\begin{figure}
    \vspace{-.25cm}
    %\hspace{-2cm}
    \includegraphics[trim={1.5cm 0 0 0}, clip, scale=.4] {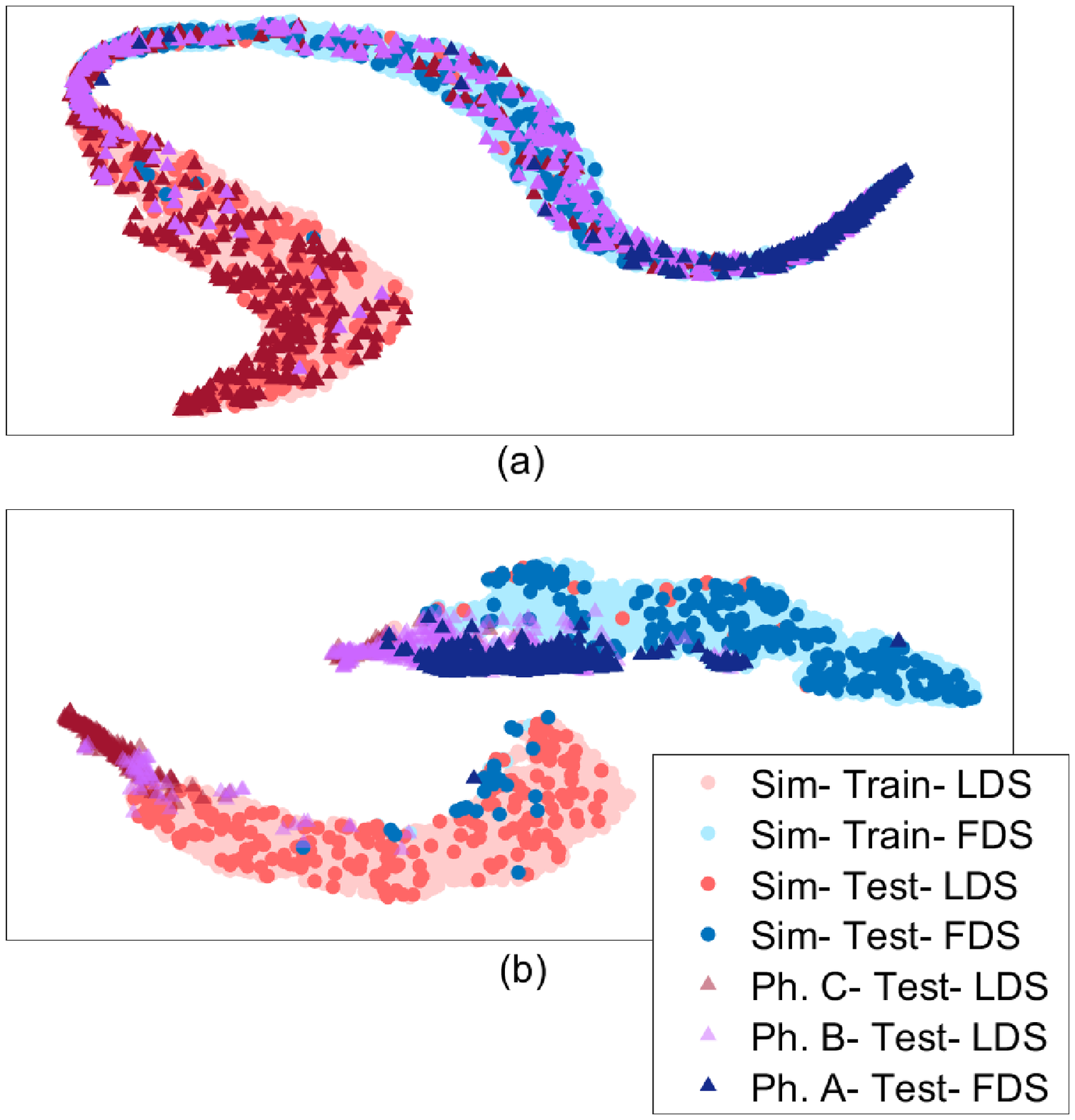}
    \vspace{-0.5cm}
    \caption{t-SNE visualization of the feature maps. (a) CNN without using patch statistics (CNN 5). (b) CNN with exploiting patch statistics (CNN 6). The feature map of CNN 6 is more separable than CNN 5. Sim: simulation data, Ph.: experimental phantom.}
    \label{figure6}
\end{figure}

To have a better intuition of the trained CNNs, we visualize the feature maps of CNN 5 and CNN 6 before the last layer using the t-SNE method \cite{maaten2008visualizing}. The feature maps are vectors of length 98 for CNN 6 and 64 for CNN 5. By the t-SNE method we reduce the dimension to be able to visualize them in 2D. Figure \ref{figure6} shows the t-SNE representation of features extracted by the CNN part of the network (Fig. \ref{figure6}A), and by the whole network (model CNN 6- Fig. \ref{figure6}B). It is clear that combining the CNN and MLP networks has led to a better separability of the two classes of LDS and FDS.\par

\begin{table*}[h]
\centering
\begin{tabular}{@{}ccccc@{}}
Model          & Input                                             &  Patch Statistics        &  \textbf{AUC (\%)}  & \textbf{Youdens' Index}\\ 
\hline
SVM & $A$                                     & \cmark                  & 0.646 (0.623- 0.669)  & 0.35  \\ 
\hline
Random forest & $A$                                     & \cmark                  & 0.895 (0.880- 0.913)  & 0.71  \\ 
\hline
MLP & $A$                                     & \cmark                  & 0.894 (0.874-0.912)  &0.72   \\ 
\hline
CNN 1              & $A$                                      & \xmark                  & 0.791 (0.781-0.802)  &0.51  \\ 
\hline
CNN 2              & $A$                                      & \cmark                  & 0.877 (0.870-0.884)&0.66  \\ 
\hline
CNN 3              & $A \times log(A)$         & \xmark         & 0.679 (0.668-0.690)         & 0.31 \\ 
\hline
CNN 4              & $A \times log(A)$        & \cmark        & 0.845 (0.837-0.853)  & 0.49         \\ 
\hline
CNN 5              & $A$, $A \times log(A)$  & \xmark        & 0.949 (0.945-0.953)           &0.79  \\ 
\hline
CNN 6              & $A$,~$A \times log(A)$  & \cmark             & 0.955 (0.951-0.959)& 0.79 \\ 
\hline
CNN 6 + ft              & $A$,~$A \times log(A)$  & \cmark             & \textbf{0.966 (0.962-0.969)}& \textbf{0.85} \\
\end{tabular}

\caption{Results of different classification models on experimental phantom data.}
\label{table3}
\end{table*}

\subsection{\textit{In vivo} Results}
\begin{figure}
    \centering
    \vspace{-.7cm}
    \hspace{-1.1 cm}
    \includegraphics[scale=.4]{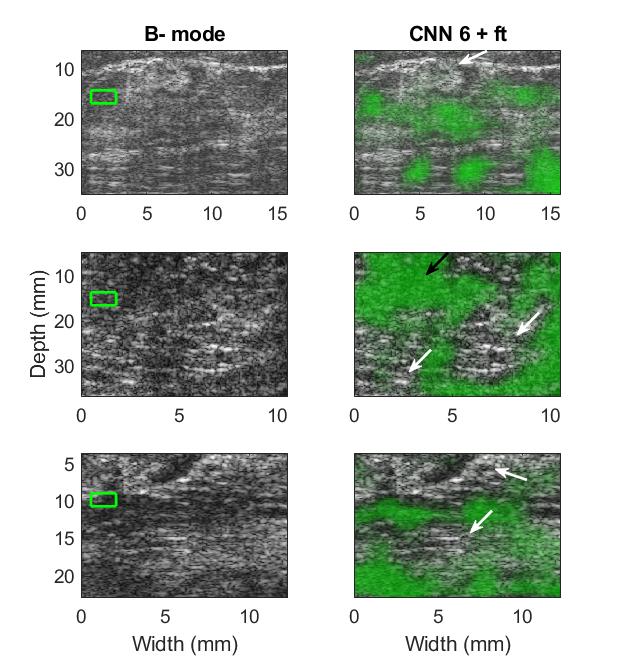}
    \caption{The results of CNN 6 + ft model on breast data from three different patients. The green overlaid color represents the areas detected as the FDS group. White arrows pinpoint the specular reflections in the ultrasound data. Black arrow shows the tissue region which looks like an FDS region in the B-mode image, as predicted by the network.}
    \label{figure9}
\end{figure}

We evaluated our proposed method using \textit{in vivo} data from three different patients. Fig. \ref{figure9} demonstrates the results of model CNN 6 + ft on the \textit{in vivo} data. The actual number of scatterers in different regions of the image is unknown. However, we expect the fat tissue in breast to be classified mostly as the FDS class, and regions with specular reflections not as FDS \cite{nasief2015acoustic}.

The patch size is 2.55 mm$\times$2.36 mm which is small to have high spatial resolution. To gain a better resolution, we considered an overlap of 87.5\% between patches when analysing the \textit{in vivo} data. As seen in the figure, the areas with specular reflections with large or aligned scatterers are detected as the LDS group, as expected and regions with relatively homogeneous speckle are classified as regions with fully developed speckle.  \par

\section{Discussion}
The density of scatterers in different parts of a tissue is an important property of that tissue which may discriminate normal and abnormal regions. Ultrasound images can be utilized to estimate this property non-invasively. This will eventually guide invasive procedures such as biopsy, leading to less expensive and safer diagnosis methods for different types of diseases. In this work, we employed DL techniques to classify the scatterer density in ultrasound images. Based on our results, different numbers of scatterers result in different texture patterns in the ultrasound image which can be detected by CNNs. Further investigations may reveal CNNs' effectiveness for other QUS problems.\par
The network trained on simulation data was able to classify the experimental phantom data, despite the fact that the number of scatterers and the imaging properties are completely different in these two datasets. Although transfer learning was not the main goal of this paper, we showed that it can help the network mitigate the domain shift problem. The impact of fine-tuning can be observed in phantom B which was challenging for the networks trained on the simulation data.   \par

%In this study, only the envelope data was used to construct and evaluate the models. Including RF data into analysis may improve the results towards the development of an automatic method for estimation of tissue properties from the ultrasound data.\par
In a fixed imaging setting, a larger number of scatterers results in a brighter ultrasound image. However, by changing the imaging machine settings, the image intensity can vary. Even though the average density of ultrasound images contains information about the scatterers concentration, it is not a reliable feature for classifying the number of scatterers, as it can be easily altered by changing the imaging setting. We eliminated the effect of the average intensity by normalizing each individual patch such that the intensity of all studied patches was in the range [0,1].\par   
The effective number of scatterers per resolution cell varies by depth and by the operating frequency. Generally, at the focal point, the resolution cell is the smallest. Therefore, there are fewer scatterers per resolution cell at the focal point compared to other regions. According to our experiments, there are more variations in real ultrasound data compared to the simulation data. One reason is that in simulation, we put the phantom with some distance to the probe, while in real ultrasound images, the entire image is obtained. Regarding the operating frequency, the small cells which are smaller than the wavelength are considered as scatterers. Therefore, the number of scatterers changes by the operating frequency (assuming the resolution cell size is fixed). In a tissue with cells in different sizes, lower center frequencies see more scatterers per resolution cell compared to the higher ones. The dependency of scatterer density on depth and frequency warrants further investigation.\par      
We included the data recorded from three phantoms in this study. The density of the scatterers is not the only parameter which differs between these phantoms. The size of the included scatterers is also different (Table \ref{table1}). However, considering the operating frequency, the size of the scatterers is still smaller than the wavelength, and does not substantially affect the results.\par
In this study, we were able to classify scatterer density in simulation and experimental phantoms. We also classified regions of homogeneous speckles as having FDS in the \textit{in-vivo} data. This information can subsequently be used to detect different types of tissue abnormalities such as tumors and cancer tissues. This study therefore provides a fundamental tool for proceeding studies on tissue characterization.\par
\section{Conclusion}
In this manuscript, we proposed different CNN models to classify small patches of ultrasound images as FDS or LDS. We proposed to use both envelope and envelope multiplied by log compressed envelope as two separate input channels to the proposed CNNs. We also benefited from patch statistics by using an MLP along with the CNN to capture information of both texture and patch statistics. We finally validated our proposed method by experimental phantoms and \textit{in vivo} data.

	% use section* for acknowledgment
%\section*{Acknowledgment}
%We acknowledge the support of the Natural Sciences and Engineering %Research Council of Canada (NSERC) RGPIN-2020-04612.

\bibliography{mybibliography}
\bibliographystyle{ieeetr}
% that's all folks
\end{document}